%% file: Comparative_tr_refl_DLA_Arxiv.tex
\documentclass[aps,prstab,superscriptaddress,longbibliography, twocolumn,amsmath,amssymb]{revtex4-2}	
\usepackage{graphicx}
\usepackage{dcolumn} 
\usepackage{hyperref}
\hypersetup{
    colorlinks=true,        
    linkcolor=blue,          
    citecolor=blue,        
    filecolor=magenta,      
    urlcolor=blue           
}
\newcommand{\Figref}[1]{Fig.\ref{#1}}

\newcommand{\citenamefont}{}
\newcommand{\bibnamefont}{}

\usepackage{color} 				
\usepackage[normalem]{ulem} 			

\begin{document}
\title{Comparative Analysis of Simulation Results of Dielectric Laser Acceleration of Non-relativistic Electrons in Transparent and Reflective Periodic Structures}
\thanks{Preprint submitted to Problems of Atomic Science and Technology November 15, 2024}

\author{I.V. Beznosenko}
\altaffiliation[Corresponding author ]{\\Email address: beznosenko1989@gmail.com (I.V. Beznosenko)}

\author{A.V.~Vasyliev}

\author{G.V.~Sotnikov}

\author{G.O.~Krivonosov}

\address{National Science Center Kharkiv Institute of Physics and Technology %
\\1, Akademichna St., Kharkiv, 61108, Ukraine}

\date{\today}

\begin{abstract}
To support of our experimental studies on dielectric laser acceleration, numerical studies of laser acceleration of nonrelativistic electrons with the initial energy of 33.9 keV in transparent and reflective periodic structures are car-ried out. On the basis of computer simulations, the acceleration rates of electrons and the quality of their beams after acceleration in compact structures of different configurations were determined and compared. Prospective acceleration schemes are proposed, in particular with reflective periodic structures, which can provide higher rates of electron acceleration in periodic structures than in previously obtained studies.
\end{abstract}
\maketitle

\section{INTRODUCTION}
Dielectric laser accelerators (DLA) attract more and more attention of researchers, first of all, due to their ultra-compactness ~\cite{England2022}.
\par The idea of dielectric laser acceleration is based on the work of R. Palmer~\cite{Palmer1980}, where it was proposed to use a periodic metallic structure irradiated by a laser pulse to accelerate electrons. However, until present basic researches on DLA was performed using dielectric structures transparent to laser radiation (chips, flat or photonic structures)~\cite{England2014}. Laser acceleration of electrons in transparent and reflective (with metal coating) periodic chip-structures in dependence of its geometric parameters is numerically investigated in~\cite{Vasyliev2021}. Numerical modeling and physical experiments on dielectric laser acceleration of non-relativistic electrons are carried out in~\cite{Breuer2013,Yousefi2019}. Different acceleration schemes are proposed in these investigations. Methods of manufacturing transparent periodic chip-structures and the influence of their geometry on acceleration rates are also described. In~\cite{Wei2018,Peralta2015}, the dielectric laser acceleration of relativistic electrons is studied in detail when the geometric parameters of periodic chip-structures are changed, including the analysis of the quality of electron beams after acceleration.
\par For the first time, the use of reflective structures was proposed in cited paper~\cite{Palmer1980}, where the acceleration of particles using reflective periodic all metallic structures (linear gratings) is described. These studies were carried out in 2D approximation (flat metallic strips). We considered an acceleration of only relativistic electron beams in paper~\cite{Vasyliev2021}.
\par Studies of the acceleration of non-relativistic electrons in reflective structures of different configurations, as well as the study of the influence of the configuration of periodic structures on the quality of non-relativistic electron beams during dielectric laser acceleration, remain relevant. Previously investigated periodic chip-structures are specific and require special production capabilities.
\par The purpose of our numerical study was to clarify the effect of changing the geometric parameters of periodic structures on the rates of laser acceleration of non-relativistic electrons and the quality of their beams, as well as conducting a comparative analysis of DLA with transparent and reflective periodic chip-structures. From a practical point of view, it was still appropriate to search for available serial periodic structures for DLA as an alternative to specific chip-structures, the manufacture of which is possible only on special equipment.
\par
\begin{figure}[!bh]
  \centering
  \includegraphics[width=0.48\textwidth]{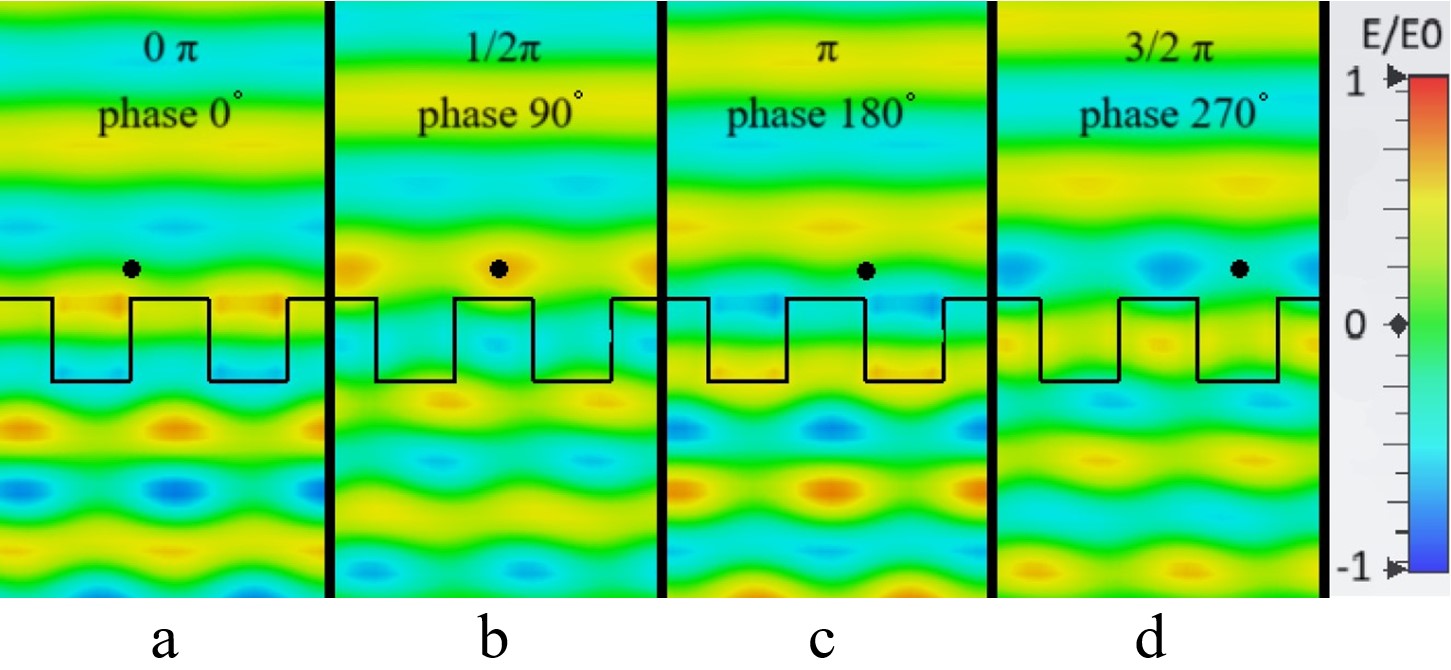}
  \caption{Schematic illustration of electron acceleration in a periodic chip-structure~\cite{Sotnikov2024}: neutral (a, c), accelerating (b), decelerating (d) phases of the field near the surface of the chip-structure. The accelerated electron is marked as a black circle.} \label{Fig:01}
\end{figure}

 Laser accelerators based on chip-structures work using the electrical component of the stimulated laser field, which is parallel to the direction of electron movement. The general principle of acceleration using a single transparent rectangular periodic chip-structure~\cite{Vasyliev2021} is demonstrated in \Figref{Fig:01}. 
\\ An electron moves in a vacuum over a periodic chip-structure from left to right. Laser radiation incidents perpendicular to the direction of its movement from below. An electric field is observed near the surface of the chip-structure, which periodically changes its direction to the opposite after every half of the laser excited wave period. Since the velocity of propagation of laser radiation in a transparent chip-structure and vacuum is different, its phase at the same moment of time is different over the center of the pillar and over the center of the groove. The greatest acceleration of the electron occurs over the center of the pillar of the chip-structure with the corresponding period, where the intensity of the electric component of the excited field is the greatest and when the direction of its action on the electron coincides with the direction of movement of this electron (\Figref{Fig:01},b). There is no acceleration above the border between the pillar and the groove (\Figref{Fig:01},a,c). Above the center of the groove, a slight slowing down of the electron is observed (\Figref{Fig:01},d). It is because the decelerating electric field, which at the time of the electron’s movement over the center of the groove has the greatest intensity over the centers of the pillars, dissipates over the groove due to diffraction effects. But this dissipated decelerating electric field is partially compensated by the accelerating field, which is formed due to difference in propagation time of laser wave in pillars and groves.
\par A change in the geometry of periodic structures leads to a change in the rates of acceleration of charged particles. Therefore, transparent and reflective periodic structures of various profiles are investigated in the pa-per to evaluate the possibility of obtaining the highest rates of acceleration of non-relativistic electrons and the most collimated beams of them after acceleration.

\section{PROBLEM FORMULATION}
\begin{figure}[!bh]
  \centering
  \includegraphics[width=0.48\textwidth]{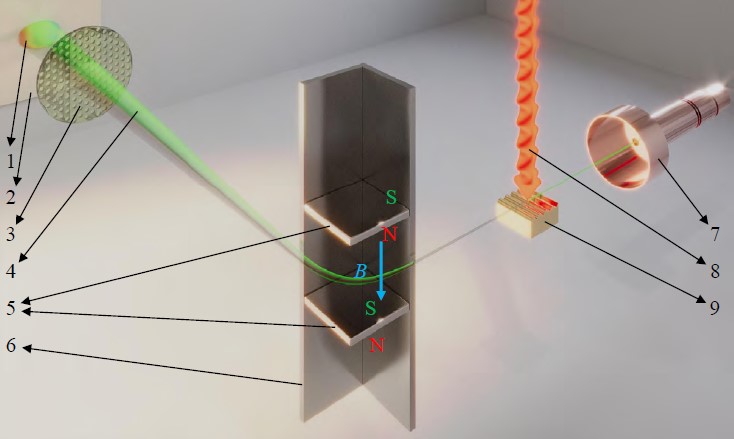}
  \caption{3D model of the DLA experiment: electrons spatially separated by energy (1) on a P43 phosphor screen (2), MCP-45-2-40-P43-CF100 microchannel plate (3), electron stream (4), magnetic spectrometer~\cite{Beznosenko2023} (N is the north pole of its magnets, S is south one, B is magnetic field direction) (5), electron gun (6), femtosecond laser beam with $\lambda$ = 800 ± $10\,nm$ (7), ThorLabs GH13-36U reflective diffraction grating as an accelerating periodic structure (scale is changed) (8).} \label{Fig:02}
\end{figure}

\Figref{Fig:02} shows the scheme of the setup for conducting our experiments on laser acceleration of electrons in order to understand the role of the studied periodic structures in DLA.
\par An electron stream with the initial energy of 33.9 keV is formed using an electron gun (6). Next, electrons are injected into a modulated periodic electromagnetic field created by a laser pulse (7), which is reflected from a reflective periodic structure (8) or passes through a transparent periodic structure~\cite{Breuer2013} (then this structure will be above the electron stream in \Figref{Fig:02}). The length of the electron stream is significantly longer than the wavelength of laser radiation, therefore, at this stage, both the acceleration of electrons that have entered the phase of the field, which accelerates them, and the deceleration of electrons that have entered the phase of the field, which slows them down, occur. Due to this, a spectrum of electrons with different energies, both accelerated and decelerated to varying degrees, will be obtained at the output after the laser accelerator. To analyze the efficiency of acceleration, it is necessary to determine their energy. For this, a magnetic spectrometer (5) will be used, which will spatially separate electrons with different energies (the most accelerated electrons are shown in red, electrons with energies close to the initial ones are shown in green, and the most decelerated ones are shown in blue). It is assumed that the number of electrons in the beam for the duration of the laser pulse will be small, therefore the spatially separated electrons (4) will pass through the MCP (3) in order to increase the output signal from the electrons (1) on the phosphor screen (2).
\par Table~\ref{Tabl_1} shows the initial parameters of the elements of the experimental setup for DLA, which we can pro-vide with the equipment available at our institute~\cite{Vasiliev2018}.

\par In order to ensure durability of periodic structures in real experiments, we use electric field amplitudes for laser acceleration that are set at one-third of the damage threshold values presented in the literature~\cite{Soong2011}, which were experimentally determined in~\cite{Breuer2013}, where the resistance of optical materials to laser-induced damage was measured. These threshold values correspond to the damage limits for the materials used in both transparent and reflective periodic structures~\cite{Wei2018}.
\begin{table}
   \centering
   \caption{Parameters of the elements of the experimental setup for dielectric laser acceleration}
   \begin{tabular}{lcc}
   \hline
   \hline
       \textbf{Parameter} & \textbf{Value}                             \\
   \hline
   \hline
          Central length of laser radiation & $\lambda$ = $800\,nm$    \\
   \hline
          Material of transparent periodic
          structures                     & $fused\,\,silica$           \\
   \hline
          Index of refraction of laser radiation                       \\
          by material of transparent periodic & $n=1.4534$             \\
          structure                                                    \\
   \hline
          Permissible amplitude of the electric field                  \\
          of laser radiation for acceleration & $E=6\,GV/m$            \\
          in a transparent periodic structure                          \\
   \hline
          Coating material of reflective periodic & $gold$             \\
          structures                                                   \\
   \hline
          Permissible amplitude of the electric field                  \\
          of laser radiation for acceleration & $E=1.8\,GV/m$          \\
          in a reflective periodic structure                           \\
   \hline
   \hline
   \end{tabular}
   \label{Tabl_1}
\end{table}

\par In order to obtain effective laser acceleration of electrons, the geometry of the periodic structure along which they move must meet the following dependencies. The period of the periodic structure must ensure the synchronism of electrons with the laser field (\Figref{Fig:03},a)~\cite{Yousefi2019}:
\begin{equation}\label{eq:01}
\begin{split}
{\lambda_p} = {\lambda}{\beta}{N},
\end{split}
\end{equation}
where $\lambda$ is the central length of laser radiation; $N=1,2,3...$ is the number of phases of one type of laser electric field above a pillar or groove (for example, when $N=1$ there is one whole accelerating (\Figref{Fig:01},b) or one whole decelerating (\Figref{Fig:01},d) phase of the field above the pillar, i.e. during the passage of the electron over the pillar, half of the wave of laser radiation passes through this electron); $\beta$ is dimensionless velocity:
\begin{equation}\label{eq:02}
\begin{split}
{\beta} = {v}/{c}=\sqrt{1-(1+{E_k}/{E_0})^{-2}},
\end{split}
\end{equation}
where $v$ is electron velocity; $c$ is velocity of laser radiation in vacuum; $E_k$ is kinetic energy of an electron; $E_0$ = 511~keV is rest energy of an electron.
\par The optimal height of the pillar (or the depth of the groove) of a transparent periodic structure ensures a change in the phase of laser radiation by $\pi$ above the pillar and the groove at the same moment in time (\Figref{Fig:03},a)~\cite{Wei2018}:
\begin{equation}\label{eq:03}
\begin{split}
{H_t} = \frac{\lambda}{2(n-1)},
\end{split}
\end{equation}
where $n$ is index of refraction of laser radiation by a material of a periodic structure.
\par The height of the pillar (or the depth of the groove) of the reflective periodic structure:
\begin{equation}\label{eq:04}
\begin{split}
{H_r} = {\lambda}/{4}.
\end{split}
\end{equation}
With this value, the electric component of the field of the laser radiation that is already reflected from the bottom of the groove will have the opposite direction to the electric component of the field that is still incident and is currently above the groove. That is, the accelerating phase of the reflected laser radiation will compensate for the decelerating phase of the incident laser radiation, into which the accelerated electron enters above the groove.

\section{SIMULATION OF TRANSPARENT PERIODIC CHIP-STRUCTURES FOR DLA}
\begin{figure}[!bh]
  \centering
  \includegraphics[width=0.48\textwidth]{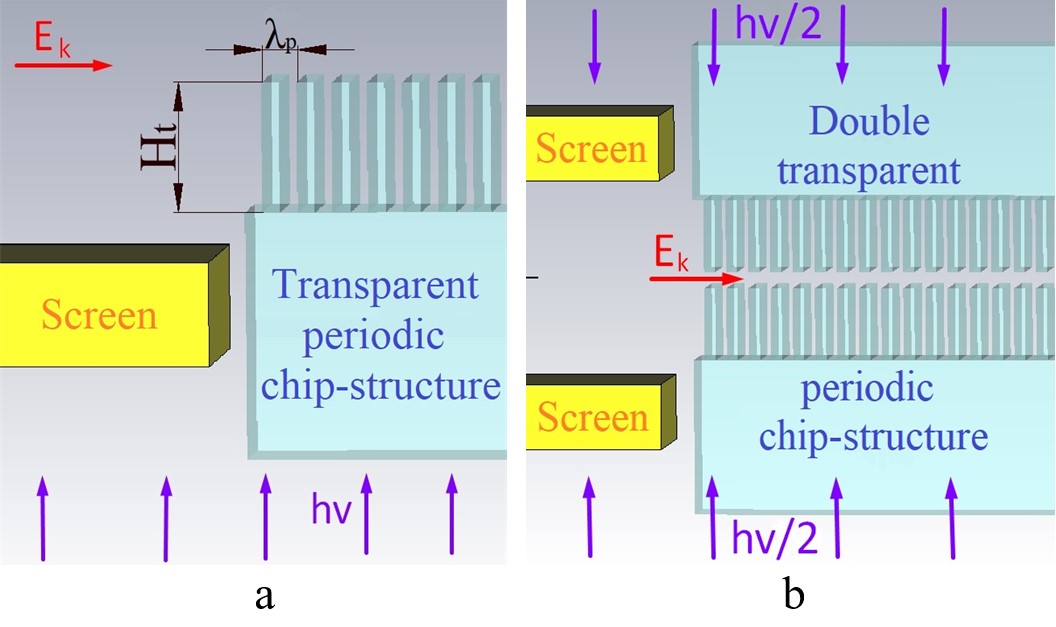}
  \caption{Single (a) and double (b) transparent periodic chip-structures for DLA ($h\nu$ is laser radiation energy).}\label{Fig:03}
\end{figure}

We performed computer simulations of electron acceleration processes with the initial energy of electrons of 33.9 keV in single (\Figref{Fig:03},a) and double (\Figref{Fig:03},b) transparent periodic chip-structures made of fused silica ($\epsilon$ = 2.112), which have a rectangular profile with a constant period of 277.769 nm, calculated by expression~(\ref{eq:01}) with $N$ = 1, by the Particle in Cell (PIC-simulation) method.
\par
\begin{figure}[!bh]
  \centering
  \includegraphics[width=0.48\textwidth]{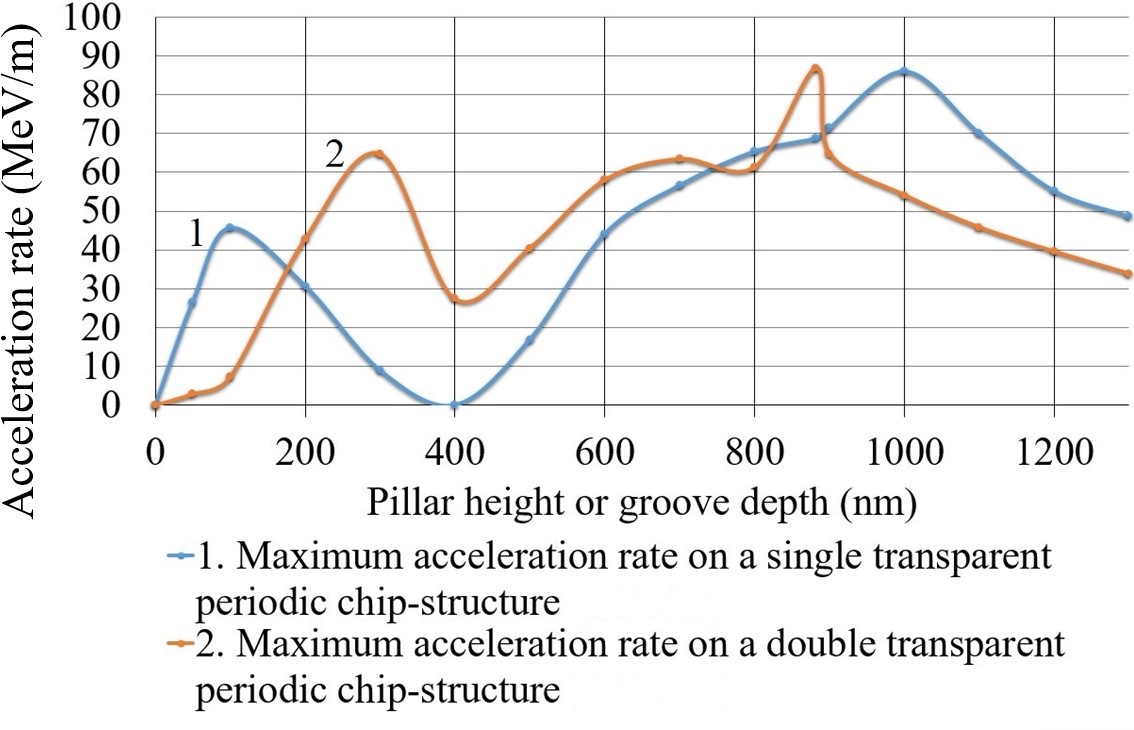}
  \caption{Dependencies of the rate of electron acceleration on the height of the pillars of a periodic chip-structure. $N$ = 1.}\label{Fig:04}
\end{figure}

\begin{figure}[!bh]
  \centering
  \includegraphics[width=0.48\textwidth]{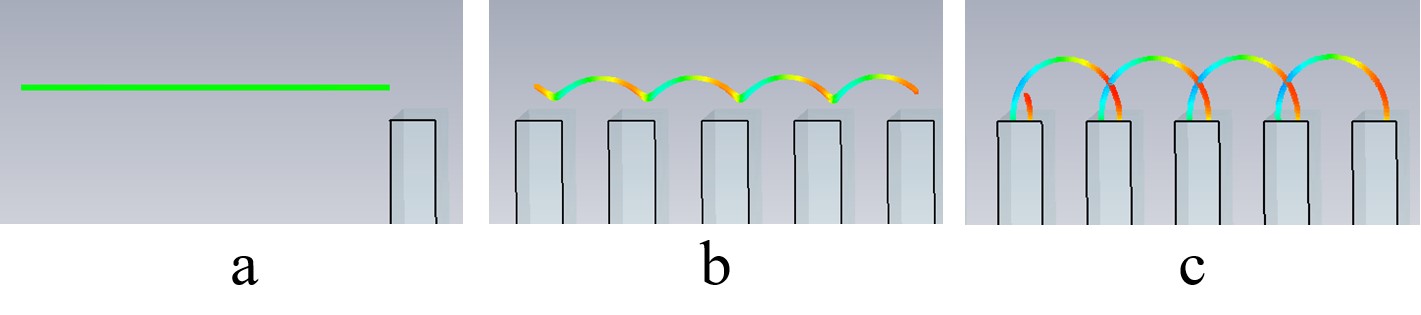}
  \caption{Spatial and energy distribution of electrons: before acceleration in a single chip-structure ($L$ = 0) (a); in the middle ($L\approx 7\,\mu m$) (b); at the end ($L\approx 14\,\mu m$) (c) of the influence of laser radiation on them. $L$ is the path that head of electron beam passed along the periodic structure, $H_t=1000\,nm$, $N = 1$.}\label{Fig:05}
\end{figure}

 The dependencies of the acceleration rate on the height of the pillars are plotted for the case when electrons move at a distance of 100 nm from the surface of the pillars (\Figref{Fig:04}).
\\ As can be seen from curve 2, expression~(\ref{eq:03}) ($H_t$ = 882~nm) is confirmed for a double chip-structure, but in practice it is difficult to obtain such a height of pillars. Therefore, structures with a pillar height of 300 nm are of interest, when an another maximum of the electron acceleration rate is observed. The same applies to a single transparent periodic chip-structure, for which the first maximum of the electron acceleration rate is ob-served at a pillar height of 100 nm (curve 1). Moreover, in the case of a single transparent periodic chip-structure, as the energy changes, the electron beam begins to diverge in the plane of laser radiation (\Figref{Fig:05}),  which leads to the fact that part of the electrons deposits on the structure, and part goes too far from the structure and stops accelerating, as shown in \Figref{Fig:05},c. Here and next, the most accelerated electrons are red, the most decelerated are blue, and green are electrons with initial energy.
\par
\begin{figure}[!bh]
  \centering
  \includegraphics[width=0.48\textwidth]{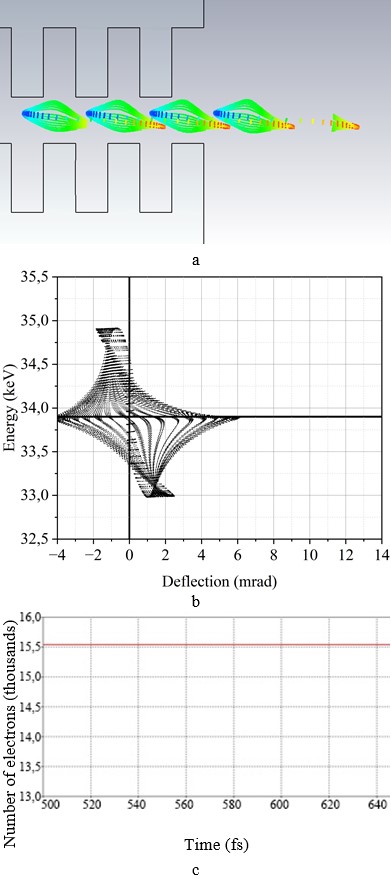}
  \caption{Spatial and energy distribution of electrons between chips of a double asymmetric periodic chip-structure (a), their deflections (b) and their presence in the space between its chips (c). $L\approx 28\,\mu m$, $H_t=300\,nm$, $N = 1$.}\label{Fig:06}
\end{figure}

\begin{figure}[!bh]
  \centering
  \includegraphics[width=0.48\textwidth]{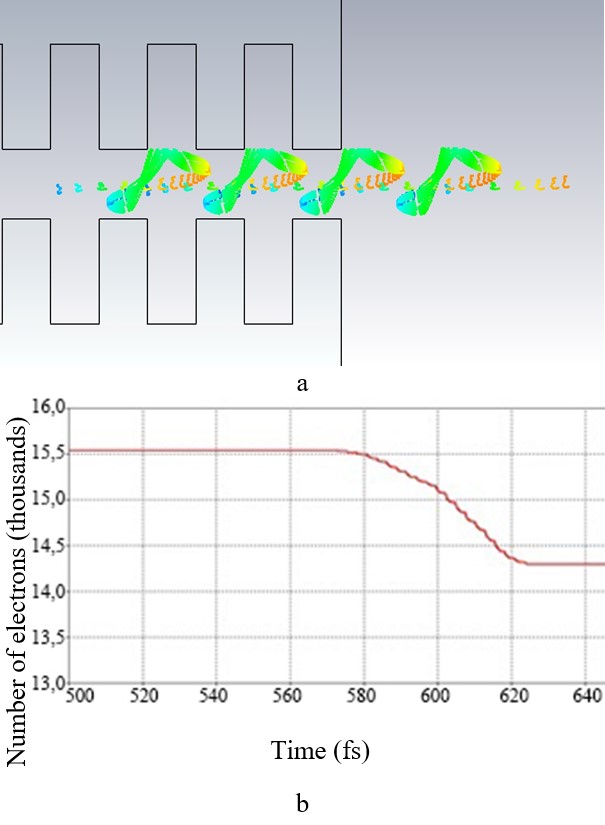}
  \caption{Spatial and energy distribution of electrons between chips of a double symmetrical periodic chip-structure (a) and their presence in the space between its chips (b). $L\approx 28\,\mu m$, $H_t=300\,nm$, $N = 1$.}\label{Fig:07}
\end{figure}

 It was also found that the divergence of electrons of the beam, whose energy was changed in the double transparent periodic chip-structure, decreases when the opposite halves of the structure are shifted relative to each other by half the period (asymmetric double periodic chip-structure) (\Figref{Fig:06},a), and fewer electrons deposit on the structure. That is, as the energy changes, part of the electrons will not deposit on the chips, as in the case of a symmetrical arrangement of the chips (symmetric double periodic chip-structure) (\Figref{Fig:07},a). The slope of the curve in the region of 600 fs in \Figref{Fig:07},b means that part of the electrons deposits on the periodic chip-structure as they diverge. The length of these periodic chip-structures was 28 $\mu m$.
Thicker lines in \Figref{Fig:06},b show the initial deflection (0 mrad) and energy (33.9 keV) of electrons. As can be seen from \Figref{Fig:06},a,b and \Figref{Fig:07},a accelerated and decelerated electrons remain more or less collimated, while electrons with energies close to the initial one diverge. It was also found that the maximum rate of electron acceleration in an asymmetric structure (G = 107 MeV/m with $H_t$ = 300 nm and G = 125 MeV/m with $H_t$ = 882~nm) is one and a half times higher than that in a symmetric structure (G = 65 MeV/m with $H_t$ = 300~nm and G = 87 MeV/m with $H_t$ = 882 nm). This predicted acceleration rate of electrons with initial energies of about 30 keV are comparable to those obtained in~\cite{Yousefi2019} (200 MeV/m), but higher than in~\cite{Breuer2013} (25 MeV/m), since the proposed configuration of periodic chip-structure are optimized according to geometric parameters. It is important to consider that if the source of laser radiation that incidents on electrons from the side of one chip is at the same geometric distance from the axis of electron movement as the source of laser radiation that incidents on electrons from the side of the opposite chip, then the initial laser radiation of such sources has be opposite in phase for the case of a symmetric double periodic chip-structure and one phase for the case of an asymmetric double periodic chip-structure when other conditions being equal.
\par Computer simulations of the acceleration processes of electrons moving at a distance of 100 nm (the acceleration rate was 53 MeV/m), 150 nm (29 MeV/m), 200~nm (10 MeV/m) and 250 nm (3 MeV/m) over a single transparent periodic chip-structure with $H_t$ = 100 nm and with $N$ = 1 (\Figref{Fig:03},a) were also carried out. The acceleration rate decreases as the distance between the electrons and the chip increases, and at a distance of more than 250 nm there is no acceleration at all, which is explained by the attenuation of laser radiation on the periodic chip-structure with a period shorter than the wavelength.
\par Finally, computer simulations of laser acceleration of electrons with the same initial energy in single transparent periodic chip-structures (\Figref{Fig:03},a) with $N$ = 1 ($\lambda_p$ = 277.769 nm, the acceleration rate was 53 MeV/m), with $N$ = 2 ($\lambda_p$ = 555.538 nm, acceleration did not occur) and with $N$ = 3 ($\lambda_p$ = 833.307 nm, the acceleration rate was 24 MeV/m) were carried out.
\par The main dependencies of electron acceleration rates on the geometry of transparent periodic chip-structures are correlated with those described in~\cite{Wei2018,Peralta2015,Wei2017}.
\section{SIMULATION OF REFLECTIVE PERIODIC STRUCTURES FOR DLA}
\begin{figure}[!bh]
  \centering
  \includegraphics[width=0.48\textwidth]{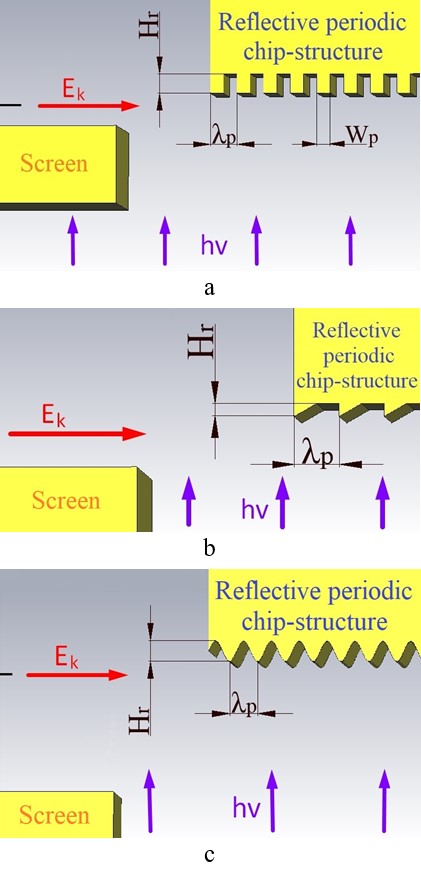}
  \caption{Single reflective periodic structures for DLA ($\lambda_p = 277,769\,nm$): with rectangular ($H_r = 200\,nm$) (a); triangular ($H_r = 80\,nm$) (b); sinusoidal geometry ($H_r = 200\,nm$) (c).}\label{Fig:08}
\end{figure}

Single periodic structures with a gold reflective coating with rectangular (\Figref{Fig:08},a), triangular (\Figref{Fig:08},b) and sinusoidal geometry (\Figref{Fig:08},c) were modeled.
\par Computer simulations of the acceleration processes of electrons with an initial energy of 33.9 keV in a beam with an initial length of approximately 1 $\mu m$ as they move along periodic structures at an initial distance of 100 nm from them were carried out. The maximum rate of electron acceleration in a periodic structure with a rectangular geometry was 125 MeV/m, with a triangular geometry was 25 MeV/m, and with a sinusoidal geometry was 37 MeV/m.
\par
\begin{figure}[!bh]
  \centering
  \includegraphics[width=0.48\textwidth]{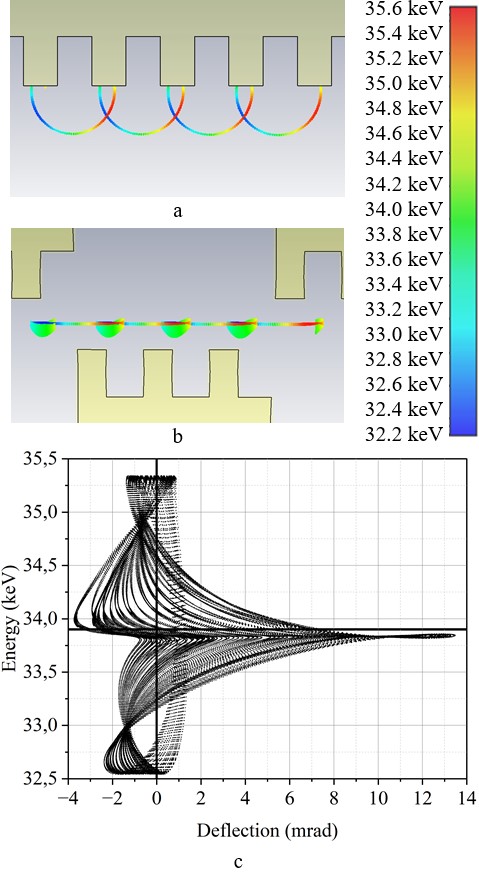}
  \caption{Spatial and energy distribution of electrons between chips of single (a), double (b) reflective periodic chip-structure and their deflections (c). $L\approx 13\,\mu m$, $H_r=200\,nm$, $N = 1$.}\label{Fig:09}
\end{figure}

\begin{figure}[!bh]
  \centering
  \includegraphics[width=0.48\textwidth]{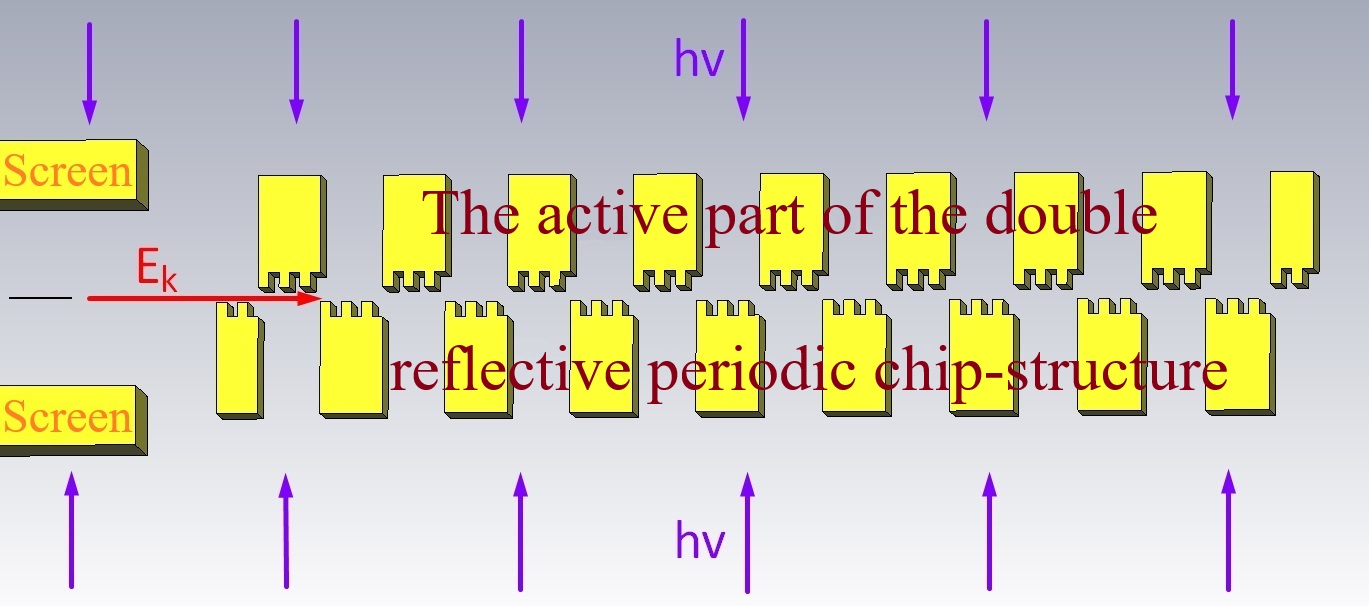}
  \caption{Part of the double reflective periodic chip-structure for DLA that affects acceleration. $\lambda_p = 277,769\,nm$, $H_r=200\,nm$, $N = 1$, $\lambda = 800\,nm$.}\label{Fig:10}
\end{figure}

 In the case of a single periodic structure, the electron beam diverges as the energy changes in the plane of the laser radiation (\Figref{Fig:09},a), which leads to the fact that part of the electrons deposits on the periodic structure, and the part moves too far from it and stops accelerate, which will be confirmed by the following simulations. To improve the quality of the electron beam, it is proposed to create such a periodic reflective chip-structure, in which laser radiation will irradiate the electron beam alternately from one side, then from the other. Part of such a double periodic chip-structure is shown in \Figref{Fig:10}, and the quality of the beam of electrons, energy of which was changed under the influence of laser radiation, is shown in \Figref{Fig:09},b,c which show that the beam remains more or less collimated during acceleration.
Thicker lines in \Figref{Fig:09},c show the initial deflection (0~mrad) and energy (33.9 keV) of electrons. It is important to consider that if the source of laser radiation that incidents on electrons from one side is at the same geometric distance from the axis of electron movement as the source of laser radiation that incidents on electrons from the opposite side, then the initial laser radiation of such sources has to be opposite in phase. The more often the alternation of the arrangement of the sections of the structure that affect acceleration occurs, first on one side, then on the other, the higher the quality of the beam. But in order for the laser radiation to pass without significant losses, the width of the gaps in the parts of the chip-structure located on one side of the axis of electron movement must be no less than the wavelength of the radiation. Therefore, in this case, the alternation occurs every three periods, since the value of the period of the chip-structure is equal to 277.769 nm, which is almost three times less than the radiation wavelength of 800 nm.
The maximum rate of electron acceleration in such a double reflective periodic chip-structure was 160~MeV/m.
\par Computer simulations of the acceleration processes of electrons moving at a distance of 100 nm (the maximum acceleration rate was 125 MeV/m), 200 nm (21 MeV/m) and 300 nm (5 MeV/m) over a single reflective periodic chip-structure with $H_r$ = 200 nm (\Figref{Fig:08},a) and $N$ = 1 (expression~(\ref{eq:01})) were also carried out. The acceleration rate decreases as the distance between the electrons and the chip increases, and at a distance of more than 300 nm, acceleration does not occur at all, which is explained by the attenuation of laser radiation on the periodic chip-structure with a period shorter than the wavelength.
\par Computer simulations of laser acceleration of electrons with the same initial energy in single reflective periodic chip-structures (\Figref{Fig:08},a) with $N$ = 1 ($\lambda_p$ = 277.769 nm, the maximum acceleration rate was 125 MeV/m), with $N$ = 2 ($\lambda_p$ = 555.538 nm, the acceleration rate was 22~MeV/m) and with $N$ = 3 ($\lambda_p$ = 833.307 nm, the acceleration rate was 47 MeV/m) were carried out.
\par
\begin{figure}[!bh]
  \centering
  \includegraphics[width=0.48\textwidth]{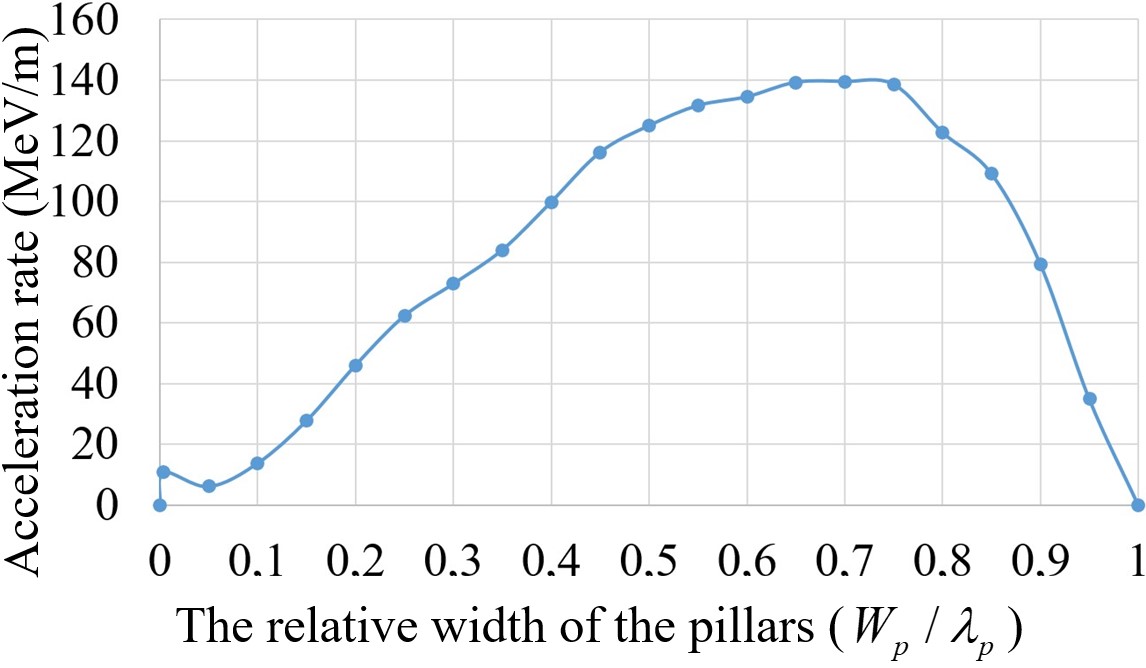}
  \caption{Dependency of the rate of laser acceleration of electrons over a single reflective periodic chip-structure on the width of its pillars. $N = 1$.}\label{Fig:11}\hfill
\end{figure}

 In all simulations of dielectric laser acceleration in periodic chip-structures with a rectangular profile presented above, the pillars in them occupied 50 \% of the period: $W_p/\lambda_p$ = 0.5 (\Figref{Fig:08},a). Computer simulations of laser acceleration of electrons with the same initial energy in single reflective periodic chip-structures with different pillar widths ($W_p$) (\Figref{Fig:11}) were carried out. As can be seen from \Figref{Fig:11}, the highest acceleration rate was when the pillars occupied 65...75 \% of the period and was approximately 139 MeV/m. A close dependence of electron acceleration efficiency on the relative width of the pillars was also observed for transparent periodic chip-structures~\cite{Wei2017}.
\par When other conditions being equal, lower acceleration rates were obtained during simulation of electron acceleration in reflective periodic chip-structures with $H_r\neq200$~nm than during simulation of electron acceleration in reflective periodic chip-structures with $H_r=200$~nm, calculated using expression~(\ref{eq:04}).
\par To fabricate the periodic chip-structures described above, nano-structure manufacturing technologies similar to the manufacturing technologies of computer microprocessors are required. For the initial experiments, it is proposed to use serial diffraction gratings with a reflective coating, which are manufactured by the ThorLabs company and which are already available at the NSC KIPT. As a result of modeling and simulation, laser electron acceleration was obtained for two reflective diffraction gratings with a sinusoidal profile (GH13-36U: acceleration rate was 30 MeV/m, GH13-18V: 27 MeV/m) and two reflective diffraction gratings with ruled (triangular) profile (GR13-1205: 16 MeV/m, GR25-1204: 14~MeV/m). The initial electron energy of 33.9 keV was chosen for our research precisely because of its correspondence to the period of these diffraction gratings according to expression~(\ref{eq:01}). The transparent diffraction gratings (GT13-12 and GTU13-12) available at our institute did not give an acceleration according to the results of computer simulation.

\section{CONCLUSIONS}

\begin{enumerate}

\item Numerical studies of the rate of laser acceleration of non-relativistic electrons and the quality of their beams after acceleration in reflective periodic chip-structures were carried out and compared with the similar studies in transparent periodic chip-structures.

\item The perspective of using reflective periodic chip-structures is shown. The use of them can provide much more efficient acceleration of electrons than the use of transparent chip-structures: the acceleration rate is one and a half times higher at three times lower intensity of laser radiation in the case of single periodic chip-structures and at one and a half times lower intensity of laser radiation in the case of double periodic chip-structures (Table~\ref{Tabl_2}). At the same time, the degree of collimation of beams of accelerated electrons is higher during acceleration in double reflective periodic chip-structure (\Figref{Fig:09},c) than in transparent one (\Figref{Fig:06},b).

\begin{table}
   \centering
   \caption{Maximum rates of laser acceleration of electrons in periodic chip-structures}
   \begin{tabular}{lcc}
   \hline
   \hline
       \textbf{A type} & \textbf{Maximum} & \textbf{\, \, \, The total}\\ \textbf{of periodic} &
       \textbf{electron} & \textbf{\, \, \, amplitude of}\\ \textbf{chip-structure} & \textbf{acceleration rate} & \textbf{\, \, \, the electric}\\ & \textbf{(initial energy} & \textbf{\, \, \, field of laser} \\
       & \textbf{is 33.9 keV)} & \textbf{\, \, \, radiation}                           \\
   \hline
   \hline
          Single transparent & $G=46$ MeV/m & \, \, \, $E=6$ GV/m\\ with $H_t=100$ nm    \\
   \hline
        Single transparent & $G=86$ MeV/m & \, \, \, $E=6$ GV/m\\ with $H_t=1000$ nm    \\
   \hline
        Single reflective & $G=125$ MeV/m & \, \, \, $E=1.8$ GV/m\\ with $H_r=200$ nm    \\
   \hline
   \hline
        Double transparent & $G=107$ MeV/m & \, \, \, $E=6$ GV/m\\ with $H_t=300$ nm    \\
   \hline
        Double transparent & $G=125$ MeV/m & \, \, \, $E=6$ GV/m\\ with $H_t=882$ nm    \\
   \hline
        Double reflective & $G=160$ MeV/m & \, \, \, $E=3.6$ GV/m\\ with $H_r=200$ nm
                                   \\
   \hline
   \hline
   \end{tabular}
   \label{Tabl_2}
\end{table}

\item The possibility of using serial reflective diffraction gratings for DLA is shown, since they can provide electron acceleration rates (30 MeV/m) sufficient for registration and practical use.

\end{enumerate}
\,

\begin{acknowledgments}
The study is supported by the National Research Foundation of Ukraine under the program “Excellent Science in Ukraine” (project \# 2023.03/0182).
\end{acknowledgments}

\input{Comparative_tr_refl_DLA_Arxiv.bbl}
\end{document}

%% file: Comparative_tr_refl_DLA_Arxiv.bbl
%

%% file: Comparative_tr_refl_DLA_Arxiv.bbl
\begin{thebibliography}{19}%
\makeatletter
\providecommand \@ifxundefined [1]{%
 \@ifx{#1\undefined}
}%
\providecommand \@ifnum [1]{%
 \ifnum #1\expandafter \@firstoftwo
 \else \expandafter \@secondoftwo
 \fi
}%
\providecommand \@ifx [1]{%
 \ifx #1\expandafter \@firstoftwo
 \else \expandafter \@secondoftwo
 \fi
}%
\providecommand \natexlab [1]{#1}%
\providecommand \enquote  [1]{``#1''}%
\providecommand \bibnamefont  [1]{#1}%
\providecommand \bibfnamefont [1]{#1}%
\providecommand \citenamefont [1]{#1}%
\providecommand \href@noop [0]{\@secondoftwo}%
\providecommand \href [0]{\begingroup \@sanitize@url \@href}%
\providecommand \@href[1]{\@@startlink{#1}\@@href}%
\providecommand \@@href[1]{\endgroup#1\@@endlink}%
\providecommand \@sanitize@url [0]{\catcode `\\12\catcode `\$12\catcode
  `\&12\catcode `\#12\catcode `\^12\catcode `\_12\catcode `\%12\relax}%
\providecommand \@@startlink[1]{}%
\providecommand \@@endlink[0]{}%
\providecommand \url  [0]{\begingroup\@sanitize@url \@url }%
\providecommand \@url [1]{\endgroup\@href {#1}{\urlprefix }}%
\providecommand \urlprefix  [0]{URL }%
\providecommand \Eprint [0]{\href }%
\providecommand \doibase [0]{https://doi.org/}%
\providecommand \selectlanguage [0]{\@gobble}%
\providecommand \bibinfo  [0]{\@secondoftwo}%
\providecommand \bibfield  [0]{\@secondoftwo}%
\providecommand \translation [1]{[#1]}%
\providecommand \BibitemOpen [0]{}%
\providecommand \bibitemStop [0]{}%
\providecommand \bibitemNoStop [0]{.\EOS\space}%
\providecommand \EOS [0]{\spacefactor3000\relax}%
\providecommand \BibitemShut  [1]{\csname bibitem#1\endcsname}%
\let\auto@bib@innerbib\@empty
\bibitem [{\citenamefont {England}\ \emph {et~al.}(2022)\citenamefont
  {England}, \citenamefont {Niedermayer}, \citenamefont {Schachter},
  \citenamefont {Hughes}, \citenamefont {Musumeci}, \citenamefont {Li},
  \ and\ \citenamefont {Kimura}}]{England2022}%
  \BibitemOpen
  \bibfield  {author} {\bibinfo {author} {\bibfnamefont {R.~J.}~\bibnamefont
  {England}}, \bibinfo {author} {\bibfnamefont {U.}~\bibnamefont {Niedermayer}},
  \bibinfo {author} {\bibfnamefont {L.}~\bibnamefont {Schachter}},
  \bibinfo {author} {\bibfnamefont {T.}~\bibnamefont {Hughes}},
  \bibinfo {author} {\bibfnamefont {P.}~\bibnamefont {Musumeci}},
  \bibinfo {author} {\bibfnamefont {R.~K.}~\bibnamefont {Li}},
  \ and\
  \bibinfo {author} {\bibfnamefont {W.~D.}~\bibnamefont {Kimura}},\
  }\bibfield  {title} {\bibinfo {title} {Considerations for a TeV collider based
  on dielectric laser accelerators},\ }\href
  {http://doi.org/10.1088/1748-0221/17/05/P05012} {\bibfield  {journal}
  {\bibinfo  {journal} {JINST}\ }\textbf {\bibinfo
  {volume} {17}},\ \bibinfo {pages} {05012} (\bibinfo {year}
  {2022})}\BibitemShut {NoStop}%
\bibitem [{\citenamefont {Palmer}(1980)}]{Palmer1980}%
  \BibitemOpen
  \bibfield  {author} {\bibinfo {author} {\bibfnamefont {R.~B.}~\bibnamefont
  {Palmer}},\ }\bibfield  {title} {\bibinfo {title} {A Laser Driven Grating Linac},\
  }\href {https://cds.cern.ch/record/1107986} {\bibfield  {journal} {\bibinfo
  {journal} {Particle Accelerators}\ }\textbf {\bibinfo {volume}
  {11}},\ \bibinfo {pages} {81--90} (\bibinfo {year} {1980})} \BibitemShut {NoStop}%
\bibitem [{\citenamefont {England}\ \emph {et~al.}(2014)\citenamefont
  {England}, \citenamefont {Noble}, \citenamefont {Bane},
  \citenamefont {Dowell}, \citenamefont {Ng}, \citenamefont {Spencer},
  \citenamefont {Tantawi}, \citenamefont {Wu}, \citenamefont {Byer},
  \citenamefont {Peralta}, \citenamefont {Soong}, \citenamefont {Chang},
  \citenamefont {Montazeri}, \citenamefont {Stephen}, \citenamefont {Wolf},
  \citenamefont {Cowan}, \citenamefont {Dawson}, \citenamefont {Gai},
  \citenamefont {Hommelhoff}, \citenamefont {Huang}, \citenamefont {Jing},
  \citenamefont {McGuinness}, \citenamefont {Palmer}, \citenamefont {Naranjo},
  \citenamefont {Rosenzweig}, \citenamefont {Travish}, \citenamefont {Mizrahi},
  \citenamefont {Schachter}, \citenamefont {Sears}, \citenamefont {Werner},
  \ and\ \citenamefont {Yoder}}]{England2014}%
  \BibitemOpen
  \bibfield  {author} {\bibinfo {author} {\bibfnamefont {R.~J.}~\bibnamefont
  {England}}, \bibinfo {author} {\bibfnamefont {R.~J.}~\bibnamefont {Noble}},
  \bibinfo {author} {\bibfnamefont {K.}~\bibnamefont {Bane}},
  \bibinfo {author} {\bibfnamefont {D.}~\bibnamefont {Dowell}},
  \bibinfo {author} {\bibfnamefont {C.-K.}~\bibnamefont {Ng}},
  \bibinfo {author} {\bibfnamefont {J.~E.}~\bibnamefont {Spencer}},
  \bibinfo {author} {\bibfnamefont {S.}~\bibnamefont {Tantawi}},
  \bibinfo {author} {\bibfnamefont {Z.}~\bibnamefont {Wu}},
  \bibinfo {author} {\bibfnamefont {R.~L.}~\bibnamefont {Byer}},
  \bibinfo {author} {\bibfnamefont {E.}~\bibnamefont {Peralta}},
  \bibinfo {author} {\bibfnamefont {K.}~\bibnamefont {Soong}},
  \bibinfo {author} {\bibfnamefont {C.-M.}~\bibnamefont {Chang}},
  \bibinfo {author} {\bibfnamefont {B.}~\bibnamefont {Montazeri}},
  \bibinfo {author} {\bibfnamefont {S.~J.}~\bibnamefont {Wolf}},
  \bibinfo {author} {\bibfnamefont {B.}~\bibnamefont {Cowan}},
  \bibinfo {author} {\bibfnamefont {J.}~\bibnamefont {Dawson}},
  \bibinfo {author} {\bibfnamefont {W.}~\bibnamefont {Gai}},
  \bibinfo {author} {\bibfnamefont {P.}~\bibnamefont {Hommelhoff}},
  \bibinfo {author} {\bibfnamefont {Y.-C.}~\bibnamefont {Huang}},
  \bibinfo {author} {\bibfnamefont {C.}~\bibnamefont {Jing}},
  \bibinfo {author} {\bibfnamefont {C.}~\bibnamefont {McGuinness}},
  \bibinfo {author} {\bibfnamefont {R.~B.}~\bibnamefont {Palmer}},
  \bibinfo {author} {\bibfnamefont {B.}~\bibnamefont {Naranjo}},
  \bibinfo {author} {\bibfnamefont {J.}~\bibnamefont {Rosenzweig}},
  \bibinfo {author} {\bibfnamefont {G.}~\bibnamefont {Travish}},
  \bibinfo {author} {\bibfnamefont {A.}~\bibnamefont {Mizrahi}},
  \bibinfo {author} {\bibfnamefont {L.}~\bibnamefont {Schachter}},
  \bibinfo {author} {\bibfnamefont {C.}~\bibnamefont {Sears}},
  \bibinfo {author} {\bibfnamefont {G.~R.}~\bibnamefont {Werner}},
  \ and\
  \bibinfo {author} {\bibfnamefont {R.~B.}~\bibnamefont {Yoder}},\
  }\bibfield  {title} {\bibinfo {title} {Dielectric laser accelerators},\ }\href
  {https://link.aps.org/doi/10.1103/RevModPhys.86.1337} {\bibfield  {journal}
  {\bibinfo  {journal} {Rev. Mod. Phys.}\ }\textbf {\bibinfo
  {volume} {86}}\bibinfo  {number} { (04)},\ \bibinfo {pages} {1337--1389}
  (\bibinfo {year} {2014})}\BibitemShut {NoStop}%
\bibitem [{\citenamefont {Vasyliev}\ \emph {et~al.}(2021)\citenamefont
  {Vasyliev}, \citenamefont {Bolshov}, \citenamefont {Galaydych},
  \citenamefont {Povrozin}, \ and\ \citenamefont {Sotnikov}}]{Vasyliev2021}%
  \BibitemOpen
  \bibfield  {author} {\bibinfo {author} {\bibfnamefont {A.~V.}~\bibnamefont
  {Vasyliev}}, \bibinfo {author} {\bibfnamefont {A.~O.}~\bibnamefont {Bolshov}},
  \bibinfo {author} {\bibfnamefont {K.~V.}~\bibnamefont {Galaydych}},
  \bibinfo {author} {\bibfnamefont {A.~I.}~\bibnamefont {Povrozin}},
  \ and\
  \bibinfo {author} {\bibfnamefont {G.~V.}~\bibnamefont {Sotnikov}},\
  }\bibfield  {title} {\bibinfo {title} {Acceleration of electron bunches using
  periodic dielectric structures with and without coating},\ }\href
  {http://doi.org/10.46813/2021-134-060} {\bibfield  {journal}
  {\bibinfo  {journal} {Problems of Atomic Science and Technology}\ }\textbf
  {\bibinfo
  {volume} {134}}\bibinfo  {number} { (04)},\ \bibinfo {pages} {60--64}
  (\bibinfo {year} {2021})}\BibitemShut {NoStop}%
\bibitem [{\citenamefont {Breuer}(2013)}]{Breuer2013}%
  \BibitemOpen
  \bibfield  {author} {\bibinfo {author} {\bibfnamefont {J.}~\bibnamefont
  {Breuer}},\ }\bibfield  {title} {\bibinfo {title} {Dielectric laser
  acceleration of non-relativistic electrons at a photonic structure},\
  }\href {https://edoc.ub.uni-muenchen.de/16147/} {\bibfield  {journal} {\bibinfo
  {journal} {Diss. Ludwig-Maximilians University (Munchen): Faculty of Physics,}\ }
  \bibinfo {pages} {10,50,ix} (\bibinfo {year} {2013})} \BibitemShut {NoStop}%
\bibitem [{\citenamefont {Yousefi}(2019)}]{Yousefi2019}%
  \BibitemOpen
  \bibfield  {author} {\bibinfo {author} {\bibfnamefont {P.}~\bibnamefont
  {Yousefi}},\ }\bibfield  {title} {\bibinfo {title} {Novel Silicon
  Nano-Structures for Dielectric Laser Accelerators},\
  }\href {https://open.fau.de/items/dcc6b397-8d07-4d6f-b576-df325bb0f6d1}
  {\bibfield  {journal} {\bibinfo
  {journal} {Diss. Friedrich-Alexander University (Erlangen-Nurnberg): Faculty
  of Natural Sciences,}\ }
  \bibinfo {pages} {13,viii} (\bibinfo {year} {2019})} \BibitemShut {NoStop}%
\bibitem [{\citenamefont {Wei}(2018)}]{Wei2018}%
  \BibitemOpen
  \bibfield  {author} {\bibinfo {author} {\bibfnamefont {Y.}~\bibnamefont
  {Wei}},\ }\bibfield  {title} {\bibinfo {title} {Investigations Into
  Dual-Grating Dielectric Laser-Driven Accelerators},\
  }\href {https://livrepository.liverpool.ac.uk/3022757/} {\bibfield  {journal}
  {\bibinfo
  {journal} {Diss. the University of Liverpool: Faculty of Science and
  Engineering,}\ }
  \bibinfo {pages} {41,15,21,46--52} (\bibinfo {year} {2018})} \BibitemShut {NoStop}%
\bibitem [{\citenamefont {Peralta}(2015)}]{Peralta2015}%
  \BibitemOpen
  \bibfield  {author} {\bibinfo {author} {\bibfnamefont {E.}~\bibnamefont
  {Peralta}},\ }\bibfield  {title} {\bibinfo {title} {Accelerator on a Chip:
  Design, Fabrication, and Demonstration of Grating-Based Dielectric
  Microstructures for Laser-Driven Acceleration of Electrons},\
  }\href {https://purl.stanford.edu/ht547xt5560}
  {\bibfield  {journal} {\bibinfo
  {journal} {Diss. Stanford University: Department of Applied Physics,}\ }
  \bibinfo {pages} {19--22} (\bibinfo {year} {2015})} \BibitemShut {NoStop}%
\bibitem [{\citenamefont {Sotnikov}\ \emph {et~al.}(2024)\citenamefont
  {Sotnikov}, \citenamefont {Vasiliev}, \citenamefont {Beznosenko},
  \citenamefont {Kovalov}, \citenamefont {Povrozin},
  \ and\ \citenamefont {Svystunov}}]{Sotnikov2024}%
  \BibitemOpen
  \bibfield  {author} {\bibinfo {author} {\bibfnamefont {G.~V.}~\bibnamefont
  {Sotnikov}}, \bibinfo {author} {\bibfnamefont {A.~V.}~\bibnamefont {Vasiliev}},
  \bibinfo {author} {\bibfnamefont {I.~V.}~\bibnamefont {Beznosenko}},
  \bibinfo {author} {\bibfnamefont {S.~M.}~\bibnamefont {Kovalov}},
  \bibinfo {author} {\bibfnamefont {A.~I.}~\bibnamefont {Povrozin}},
  \ and\
  \bibinfo {author} {\bibfnamefont {O.~O.}~\bibnamefont {Svystunov}},\
  }\bibfield  {title} {\bibinfo {title} {Comparative Analysis of Electron
  Acceleration by Laser Pulse in Flat and Chip Dielectric Structures},\ }\href
  {https://doi.org/10.48550/arXiv.2409.19313} {\bibfield  {journal}
  {\bibinfo  {journal} {arXiv:2409.19313 [physics.acc-ph],}\ } \bibinfo {pages} {12}
  (\bibinfo {year} {2024})}\BibitemShut {NoStop}%
\bibitem [{\citenamefont {Beznosenko}\ \emph {et~al.}(2023)\citenamefont
  {Beznosenko}, \citenamefont {Vasyliev}, \ and\ \citenamefont
  {Sotnikov}}]{Beznosenko2023}%
  \BibitemOpen
  \bibfield  {author} {\bibinfo {author} {\bibfnamefont {I.~V.}~\bibnamefont
  {Beznosenko}}, \bibinfo {author} {\bibfnamefont {A.~V.}~\bibnamefont {Vasyliev}},
  \ and\
  \bibinfo {author} {\bibfnamefont {G.~V.}~\bibnamefont {Sotnikov}},\
  }\bibfield  {title} {\bibinfo {title} {Simulation of deflecting system based
  on permanent magnets with a non-uniform magnetic field to register accelerated
  and decelerated electrons generated in dielectric laser accelerator},\ }\href
  {http://doi.org/10.46813/2023-148-110} {\bibfield  {journal}
  {\bibinfo  {journal} {Problems of Atomic Science and Technology}\ }\textbf
  {\bibinfo
  {volume} {148}}\bibinfo  {number} { (06)},\ \bibinfo {pages} {110--115}
  (\bibinfo {year} {2023})}\BibitemShut {NoStop}%
\bibitem [{\citenamefont {Vasiliev}\ \emph {et~al.}(2018)\citenamefont
  {Vasiliev}, \citenamefont {Dovbnya}, \citenamefont {Yegorov},
  \citenamefont {Zaitsev}, \citenamefont {Leshchenko}, \citenamefont {Onischenko},
  \citenamefont {Povrozin}, \ and\ \citenamefont {Sotnikov}}]{Vasiliev2018}%
  \BibitemOpen
  \bibfield  {author} {\bibinfo {author} {\bibfnamefont {A.~V.}~\bibnamefont
  {Vasiliev}}, \bibinfo {author} {\bibfnamefont {A.~N.}~\bibnamefont {Dovbnya}},
  \bibinfo {author} {\bibfnamefont {A.~M.}~\bibnamefont {Yegorov}},
  \bibinfo {author} {\bibfnamefont {V.~P.}~\bibnamefont {Zaitsev}},
  \bibinfo {author} {\bibfnamefont {V.~P.}~\bibnamefont {Leshchenko}},
  \bibinfo {author} {\bibfnamefont {I.~N.}~\bibnamefont {Onischenko}},
  \bibinfo {author} {\bibfnamefont {A.~I.}~\bibnamefont {Povrozin}},
  \ and\
  \bibinfo {author} {\bibfnamefont {G.~V.}~\bibnamefont {Sotnikov}},\
  }\bibfield  {title} {\bibinfo {title} {Works in the NSC KIPT on the creation
  and application of the CPA laser system},\ }\href
  {https://vant.kipt.kharkov.ua/ARTICLE/VANT_2018_4/article_2018_4_289.pdf}
  {\bibfield  {journal}
  {\bibinfo  {journal} {Problems of Atomic Science and Technology}\ }\textbf
  {\bibinfo
  {volume} {116}}\bibinfo  {number} { (04)},\ \bibinfo {pages} {289--293}
  (\bibinfo {year} {2018})}\BibitemShut {NoStop}%
\bibitem [{\citenamefont {Soong}\ \emph {et~al.}(2011)\citenamefont
  {Soong}, \citenamefont {Byer}, \citenamefont {McGuinness},\citenamefont
  {Peralta},\ and\ \citenamefont {Colby}}]{Soong2011}%
  \BibitemOpen
  \bibfield  {author} {\bibinfo {author} {\bibfnamefont {K.}\ \bibnamefont
  {Soong}}, \bibinfo {author} {\bibfnamefont {R.~L.}\ \bibnamefont {Byer}},\
  \bibinfo {author} {\bibfnamefont {C.}\ \bibnamefont {McGuinness}},
  \bibinfo {author} {\bibfnamefont {E.}\ \bibnamefont {Peralta}},
  and\ \bibinfo {author} {\bibfnamefont {E.}~\bibnamefont
  {Colby}},\ }\bibfield  {title} {\bibinfo {title} {Experimental determination
  of damage threshold characteristics of IR compatible optical materials},\ }in\
  \href {https://accelconf.web.cern.ch/pac2011/papers/mop095.pdf} {\emph
  {\bibinfo {booktitle} {Proceedings of 2011 Particle Accelerator
  Conference}}, \bibinfo  {publisher} {PAC’11 OC/IEEE, New York, USA},\
  \bibinfo {pages} {277--279} (\bibinfo {year} {2011})}\BibitemShut {NoStop}%
\bibitem [{\citenamefont {Wei}\ \emph {et~al.}(2017)\citenamefont
  {Wei}, \citenamefont {Jamison}, \citenamefont {Xia},
  \citenamefont {Hanahoe}, \citenamefont {Li}, \citenamefont {Smith},
  \ and\ \citenamefont {Welsch}}]{Wei2017}%
  \BibitemOpen
  \bibfield  {author} {\bibinfo {author} {\bibfnamefont {Y.}~\bibnamefont
  {Wei}}, \bibinfo {author} {\bibfnamefont {S.}~\bibnamefont {Jamison}},
  \bibinfo {author} {\bibfnamefont {G.}~\bibnamefont {Xia}},
  \bibinfo {author} {\bibfnamefont {K.}~\bibnamefont {Hanahoe}},
  \bibinfo {author} {\bibfnamefont {Y.}~\bibnamefont {Li}},
  \bibinfo {author} {\bibfnamefont {J.~D.~A.}~\bibnamefont {Smith}},
  \ and\
  \bibinfo {author} {\bibfnamefont {C.~P.}~\bibnamefont {Welsch}},\
  }\bibfield  {title} {\bibinfo {title} {Beam quality study for a grating-based
  dielectric laser-driven accelerator},\ }\href
  {http://dx.doi.org/10.1063/1.4975080}
  {\bibfield  {journal}
  {\bibinfo  {journal} {Physics of Plasmas}\ }\textbf
  {\bibinfo
  {volume} {24}}\bibinfo  {number} { (02)},\ \bibinfo {pages} {023102-3}
  (\bibinfo {year} {2017})}\BibitemShut {NoStop}%
\end{thebibliography}
